%% file: main.tex
\documentclass[sigconf]{acmart}

\AtBeginDocument{%
  \providecommand\BibTeX{{%
    \normalfont B\kern-0.5em{\scshape i\kern-0.25em b}\kern-0.8em\TeX}}}

\copyrightyear{2026} 
\acmYear{2026} 
\setcopyright{rightsretained} 

\usepackage{amsmath}
\usepackage{xspace}
\usepackage{balance}
\usepackage{enumitem}
\usepackage{ifthen}
\usepackage{subcaption}
\usepackage{hyperref}
\usepackage{multirow}
\usepackage{graphicx}
\usepackage{framed}
\usepackage{microtype}
\usepackage{xcolor}
\usepackage{balance}
\usepackage{listings}
\usepackage{multirow}

\copyrightyear{2026}
\acmYear{2026}
\setcopyright{cc}
\setcctype{by}
\acmConference[FSE Companion '26]{34th ACM Joint European Software Engineering Conference and Symposium on the Foundations of Software Engineering}{July 05--09, 2026}{Montreal, QC, Canada}
\acmBooktitle{34th ACM Joint European Software Engineering Conference and Symposium on the Foundations of Software Engineering (FSE Companion '26), July 05--09, 2026, Montreal, QC, Canada}

\lstdefinestyle{mystyle}{
  backgroundcolor=\color{white},commentstyle=\color{green},
  keywordstyle=\color{purple},
  numberstyle=\tiny\color{gray},
  stringstyle=\color{purple},
  basicstyle= \small,
  breakatwhitespace=false,
  breaklines=true,
  captionpos=b,
  keepspaces=true,
  numbers=left,
  numbersep=5pt,
  showspaces=false,
  showstringspaces=false,
  showtabs=false,
  tabsize=2,
}
{\endMakeFramed}

\newboolean{showcomments}
\setboolean{showcomments}{true}

\ifthenelse{\boolean{showcomments}}
{\newcommand{\nb}[2]{
		\fbox{\bfseries\sffamily\scriptsize#1}
		{\sf\small$\blacktriangleright$\textit{#2}$\blacktriangleleft$}
	}
}
{\newcommand{\nb}[2]{}
}

\newcommand\ZOHAIB[1]{\textcolor{blue}{\nb{ZOHAIB}{#1}}}
\newcommand\DANIELE[1]{\textcolor{blue}{\nb{DANIELE}{#1}}}

\newcommand{\ie}{\textit{i.e.,}\xspace}
\newcommand{\eg}{\textit{e.g.,}\xspace}

\newcommand{\etal}{\textit{et al.}\xspace}
\newcommand{\figref}[1]{Fig.~\ref{#1}\xspace}
\newcommand{\tabref}[1]{Table~\ref{#1}\xspace}

\newcommand{\rqone}{How are ML models used in the studied open-source projects to make decisions?}
\newcommand{\rqthree}{What kinds of post-processing do projects perform on ML-based decisions?}
\newcommand{\rqfour}{To what extent do models' usage comply with terms of use and other regulations?}

\acmSubmissionID{}

\begin{document}
\sloppy

\title[Machine Learning in the Wild]{Machine Learning in the Wild: Early Evidence of Non-Compliant ML-Automation in Open-Source Software}

\settopmatter{authorsperrow=4}

\author{Zohaib Arshid}
\affiliation{%
 \institution{University of Sannio}\city{Benevento}\country{Italy}
}
\email{z.arshid@studenti.unisannio.it}

\author{Daniele Bifolco}
\affiliation{%
 \institution{University of Sannio}\city{Benevento}\country{Italy}
}
\email{d.bifolco@studenti.unisannio.it}

\author{Fiorella Zampetti}
\affiliation{%
  \institution{University of Sannio}
  \city{Benevento}
  \country{Italy}
}
\email{fzampetti@unisannio.it}

\author{Massimiliano Di Penta}
\affiliation{%
 \institution{University of Sannio}\city{Benevento}\country{Italy}
}
\email{dipenta@unisannio.it}

\renewcommand{\shortauthors}{Arshid, Bifolco, Zampetti, and Di Penta}

\begin{abstract} 
The increasing availability of Machine Learning (ML) models, particularly foundation models, enables their use across a range of downstream applications, from scenarios with missing data to safety-critical contexts. This, in principle, may contravene not only the models' terms of use, but also governmental principles and regulations.
This paper presents a preliminary investigation into the use of ML models by 173 open-source projects on GitHub, spanning 16 application domains.  We evaluate whether models are used to make decisions, the scope of these decisions, and whether any post-processing measures are taken to reduce the risks inherent in fully autonomous systems. Lastly, we investigate the models' compliance with established terms of use. This study lays the groundwork for defining guidelines for developers and creating analysis tools that automatically identify potential regulatory violations in the use of ML models in software systems. 
\end{abstract}



\maketitle

\section{Introduction}
\label{sec:intro}
\input{intro}

\section{Study Design} 
\label{sec:design}

\input{design}

\section{Study Results}
\label{sec:results}
\input{results}

\section{Threats To Validity}
\label{sec:threats}

\input{threats}


\section{Conclusion} 
\label{sec:conclusion}
\input{conclusion}


\begin{acks} 
This work is supported by the InnoGuard Marie Skłodowska-Curie Doctoral Network (Grant Agreement No. 101169233). Daniele Bifolco is supported by the PhD Scholarship funded by DM 118/2023.
\end{acks}
\balance

\bibliographystyle{ACM-Reference-Format}
\bibliography{bibliography}


\end{document}

%% file: intro.tex
Machine Learning (ML) models are widely used to support automation in several application domains. In ``low-risk", highly human-assisted domains, such as productivity and entertainment, ML models can be seamlessly integrated into applications with minimal friction. Conversely, in mission- or safety-critical domains, such as autonomous vehicles and healthcare, where ML models can be used as the primary decision-maker, their integration requires far greater scrutiny. 
Also, leveraging ML for mission- and safety-critical tasks is subject to governmental regulations. For instance, the EU AI Act~\citep{aiact2024} requires full transparency for general-purpose AI foundation models. Specifically, ANNEX III~\cite{annexIIIeuaiact_2026} defines the \emph{areas} of \textit{High-Risk AI Systems} (in the following referred to as ``high-risk ML projects''), identifying applications that are either subject to strict regulatory oversight or deemed entirely unacceptable: 
\begin{quote}
\emph{
    This Annex lists use cases that would qualify an AI system as ``high-risk'' [..] include AI used in biometrics, critical infrastructure, education, employment, essential services, law enforcement, migration, justice [..] and financial fraud.}
\end{quote}

This raises the need to answer the following question: \emph{\textbf{To what extent can software systems autonomously execute safety-critical tasks via ML, and what countermeasures are in place to manage the risks of a fully model-driven decision-making?}}

Prior studies have started exploring software engineering practices when accounting for regulation, such as the traceability between regulatory requirements and source code~\cite{Cleland-HuangCGE10}, the compliance of software systems to laws and regulations~\cite{MarczakCzajkaDC25}, and the generation of software documentation~\cite{SovranoHAB25} to support compliance verification with respect to the EU AI Act. 
Dearstyne \etal ~\cite{DearstyneGCC25} have defined a framework to evaluate the trustworthiness and safety of AI techniques, \eg reinforcement learning, within cyber-physical systems.
Scantamburlo \etal \cite{3643691.3648589} conducted a student challenge in which they identified key issues in addressing system compliance with the EU AI Act, finding the need   
to define a suitable regulatory framework for traceability.

To the best of our knowledge, there is no specific work examining the use of ML models in high-risk projects and their compliance with underlying regulations. This paper discusses the results of a preliminary investigation on the use of ML models in high-risk ML projects in the open-source ecosystem. Specifically, we started by identifying open-source projects hosted on GitHub across 16 application domains. In total, we obtained {173} projects that leverage 
ML models, representing the core objects of our analysis.

We subsequently analyze these projects to (i) understand whether the projects leverage ML models to make decisions and also to characterize the types of decisions taken by the ML models, (ii) determine the post-processing safeguards, if any, applied on top of the automated decision, and (iii) evaluate the violations of the models' usage with respect to their Terms of Use (ToU).  

This study lays the groundwork for defining guidelines to support developers in creating and evolving legally compliant ML-intensive systems, as well as for developing approaches and tools to automatically verify deviations from these requirements.

The replication package is available online~\cite{replication}.

%% file: design.tex
The \emph{goal} of the study is to analyze automated decision-making processes in open-source, high-risk ML projects.
The \emph{quality focus} is on the trustworthiness of these models and their adherence to legal and regulatory frameworks.
The \emph{perspective} is that of software developers who create systems that leverage ML to automate decision-making, as well as of practitioners who may violate models' ToU and regulations when relying on ML models. 
The \emph{context} comprises {173} high-risk open-source Python projects from 16 application domains that use at least one ML model. 
To reach our goal, the study answers the following three research questions:
\\
\begin{itemize}
    \item \textbf{RQ$_1$:} \emph{\rqone}
    \item \textbf{RQ$_2$:} \emph{\rqthree}
    \item \textbf{RQ$_3$:} \emph{\rqfour}
\end{itemize}


\subsection{Study Context}
\label{sec:context}

The project selection procedure is based on GitHub topics, \ie labels used by developers to categorize their projects, making them highly discoverable. 
By analyzing the EU AI Act's descriptions of high-risk AI systems~\cite{annexIIIeuaiact_2026}, we identified 16 keywords (listed in the first column of \tabref{tab:github_projects}) that map to risky application areas. 
After that, we leveraged the GitHub Search API to search for topics related to any of our keywords, \ie \url{https://github.com/search?q="keyword"&type=topics}. For each keyword (hereafter referred to as ``domain''), GitHub retrieves topics related to it by performing string matches against topic names and descriptions, and returns a ranked list based on volume, activity, and community engagement. To ensure relevance and popularity, we limited our selection to topics that appeared on the first page. 
For each topic on this page, we retrieved all repositories (later referred to as ``projects'') that referenced it, and kept only those that used Python as the main programming language. Furthermore, we discarded niche topics by excluding those with fewer than five projects using them.
We retrieved a total of 12,523 projects, distributed across 16 domains as shown in \tabref{tab:github_projects}. To ensure our analysis focused exclusively on projects that rely on ML models, we intersected this list with the dependents of the \textit{huggingface/transformers} library. Note that this may exclude ML projects that use models with other libraries, despite the widespread use of the \emph{transformers} library. Such a choice also allowed us to focus on a fairly limited set of projects, given that the rest of the analysis is mostly manual.
This left us with 173 GitHub projects manually analyzed to answer our research questions.

\subsection{Analysis Methodology}
\label{sec:prj_analysis}

To answer \textbf{RQ$_1$}, we manually reviewed each project to 
characterize the role of the ML models being used. Specifically, we determine whether a project has a fully automated decision-making mechanism or requires human oversight (human-in-the-loop) to validate or finalize decisions. We did this by:


\begin{enumerate}
\item \textbf{Looking at documentation, \ie \texttt{README}:} to determine the project's purpose and the role played by the ML model; 
\item \textbf{Locating the decision point in the source code:} to determine the extent to which the project relies on a fully automated decision-making process, \ie the ML model's output directly influences the program's behavior, or else the ML model output is subject to a human interpretation.
\end{enumerate}

To locate the decision point, we first identified model usage by scanning Python source code files for framework-specific imports, \eg \texttt{from langchain\_openai import ChatOpenAI}. Using those imports, we traced the projects' configuration files and source code to locate the model instantiation and its defined parameters. 
Subsequently, we identified the inference invocation points  by looking at (i) high-level HF API, \eg  \texttt{pipeline}, \texttt{generate}, or (ii) framework-specific execution functions in PyTorch, TensorFlow, and ONNX, \eg \texttt{predict}, \texttt{forward}. 
After that, we performed a control-flow analysis from the inference point to determine how model outputs are consumed and to what extent they influence the system behavior. 
The manual annotation procedure has been applied to the initial set of 173 projects to discard projects that do not directly use the model output in condition logic, control flow, or automated actions that do not require mandatory human intervention. 
Once we identified a set of projects in which ML models directly drive decision-making, using a card sorting procedure \cite{cardSorting}, we manually classified---with an author performing the analysis and a second author conducting a validation check---the nature of these decisions based on two primary factors: the role of the ML model within the project and the use made of its output in terms of system behavior. 

To answer \textbf{RQ$_2$}, we manually inspected the source code of each project to understand how the model outputs are transformed into decisions and identify common post-processing patterns, \ie data conversion and guards aimed at checking the validity of the ML model output. We did that because we wanted to determine whether the project performs any further action---deterministic or not---to mitigate the risks of fully relying on the ML model.
To do so, we examined the structural characteristics of the post-processing logic applied to the model output. These include, for example, the use of confidence threshold or probability score, binary decision rules, regular expression-based extraction, schema enforcement, hierarchical or multi-stage validation checks, domain-specific rules such as clinical protocols, and aggregation mechanisms, \eg voting or averaging.
Note that a project can use multiple ML models to make distinct decisions, and a single model output can undergo various post-processing steps. For these reasons, we analyzed each decision separately.
Starting from an initial set of observed post-processing actions, we grouped projects based on shared structural characteristics of their post-processing logic. To derive a qualitative code, we conduct an iterative cluster refinement procedure. 

To answer \textbf{RQ$_3$}, we investigate whether the use of the ML model by the project is compliant with existing regulations. To do this, we retrieved the ToU documentation either by the model's owner or from the hosting platform, \ie Hugging Face or GitHub. 
Next, we assessed whether the project's context, decision-making behavior, and post-processing of model output comply with the ToU for the ML model. Our focus was on identifying violations related to application domains, prohibited use cases, and human-in-the-loop requirements. If multiple models were used, we analyzed each one separately. 

%% file: results.tex
\begin{table}[t]
\centering
\caption{Studied systems making automated decisions}
\label{tab:github_projects}
\small 
\resizebox{\columnwidth}{!}{
    \begin{tabular} {lrrr}
        \toprule
        \multirow{2}{*}{\textbf{Keywords}} & \textbf{\# GitHub}  & {\textbf{\# Systems Using}} & {\textbf{\# Decision-making}} \\
         & \textbf{Systems} &  \textbf{Hugging Face Models} & \textbf{Systems} \\ 
        \midrule
        Healthcare & 1,215 & 50 & 16 \\
        Autonomous Driving & 1,847 & 19 & 9 \\
        Finance & 3,086 & 50 & 5 \\
        Blockchain & 1,345 & 15 & 4 \\
        Cybersecurity & 1,377 & 17 & 4 \\
        Drones & 777 & 3 & 3 \\
        Agriculture & 334 & 3 & 1 \\
        Compliance-Legal & 267 & 4 & 1 \\
        Energy & 920 & 8 & 1 \\
        Telecommunication & 80 & 1 & 1 \\
        Banking & 503 & 1 & 0 \\
        Transportation & 399 & 0 & 0 \\
        Automotive & 179 & 0 & 0 \\
        Smart-city & 81 & 1 & 0 \\
        Automated-hiring & 93 & 1 & 0 \\
        Dynamic-pricing & 20 & 0 & 0 \\
        \midrule
        \textbf{Total} & \textbf{12,523} & \textbf{173} & \textbf{45} \\
        \bottomrule
    \end{tabular}
    }
    \vspace{-3mm}
\end{table}

\textbf{RQ$_1$:} \textbf{\rqone} 

As shown in \tabref{tab:github_projects}, out of 173 high-risk ML projects, 45 (26\%) employ ML models for automated decision-making, with the most frequent applications found in healthcare (16) and autonomous driving (9). Regarding the level of automation, {36} projects implement a \textbf{Fully Automated} decision, whereas 9 use a \textbf{Partially Automated} decision involving a human-in-the-loop. 
For example, in \texttt{s-suryakiran/DriveVLM}, the \texttt{LLVM\_Agent.py} file takes sensor data, sends it to a vision language model, and uses its output to apply throttle and brake commands directly, without any human approval step.
Instead, \texttt{KStar1014/Medical-Assistant-Multi-Agents} employs a human-in-the-loop check for all medical image diagnoses. Specifically, the model predicts a medical condition by analyzing an image, which is then further evaluated by a human to finalize the diagnosis. In this process, the \texttt{needs\_human\_validation} flag is set to \texttt{True} whenever critical predictions require professional approval. This action pauses the workflow and prompts healthcare experts to confirm or reject the model's decision.

As reported in \tabref{tab:output_clusters}, we classified the decisions made by the analyzed projects into nine categories. Note that one project (\emph{NguyenHuy190303/Salus-Analytica}) belongs to two categories (Activity/State Classification and Therapeutic Generation).
The most prevalent category, \textbf{AI Recommender}, includes 20 projects that generate recommendations to directly inform decision-making. 
Note that, while providing recommendations may not always be classified as ``automated decision'', we included them because such advisories can lead to serious consequences, particularly in high-risk domains.
A representative example is the finance project \texttt{Andrechang/Atradebot}, which automates buy/sell decisions and portfolio advice.
\textbf{Automated Control} includes 7 projects whose models output 
control signals for physical systems. For instance, \texttt{autonomousvision/plant} uses multi-sensor perception and a transformer-based planner to generate driving paths and control commands for self-driving vehicles.
\textbf{Threat Detection} features 4 projects focused on identifying security threats or risks. An example is \texttt{cprite/phishing-detection-ext}, which classifies text inputs as phishing or not.
\textbf{Visual Generation \& Simulation} projects (3) generate or simulate visual environments, \eg \texttt{OpenDriveLab/Vista}, that produces synthetic driving scenes. 
\textbf{Decision Optimization} (3) projects output optimized operational decisions, \eg \texttt{AppSolves/LanePilot} recommends lane allocations to improve traffic flow. 
\textbf{Governance \& Compliance} projects (3) generate outputs (\eg approvals, compliance scores, or audit reports) that are used to assess whether project behaviors conform to predefined policies or regulatory requirements.
Finally, \textbf{Visual Perception, Activity/State Classification, and Therapy Generation} are the smallest categories, each with 2 projects. These include projects that interpret visual inputs (\eg \texttt{bowang-lab/MedRAX}), classify human or system states (\eg \texttt{Anish15AG/mental-health-prediction}), or generate adaptive therapeutic content (\eg \texttt{theinterneti/TTA}).
\\

\begin{table}[t]
\centering
\small
\caption{Automated decision types distribution}
\label{tab:output_clusters}
\setlength{\tabcolsep}{4pt}
\begin{tabular}{lr}
\toprule
\textbf{Category Name} & \textbf{\# Systems} \\
\midrule
AI Recommender & 20 \\
Automated Control & 7 \\
Threat Detection & 4 \\
Visual Generation \& Simulation & 3 \\
Decision Optimization & 3 \\
Governance \& Compliance & 3 \\
Visual Perception & 2 \\
Activity/State Classification & 2 \\
Therapy Generation & 2 \\
\bottomrule
\end{tabular}
\vspace{-1mm}
\end{table}

\begin{table}[t]
\centering
\caption{Post-Processing categories distribution}
\label{tab:postprocessing}
\small
\begin{tabular}{lr}
\toprule
\textbf{Post-Processing Categories} & \textbf{Frequency} \\
\midrule
Output Structuring & 13 \\
Output Consistency & 11 \\
Auto-Refiner & 9 \\
Multi-Agent Coordination & 9 \\
Confidence Threshold & 7 \\
\bottomrule
\end{tabular}
\vspace{-3mm}
\end{table}

\textbf{RQ$_2$:} \textbf{\rqthree}



Out of 45 studied projects, 40 apply at least one post-processing step to the output of the ML model, while 5 use the raw output to decide without further processing. In other words, most of the studied projects do not depend directly on
raw ML outputs but perform additional steps before using them.
\tabref{tab:postprocessing} reports the types of post-processing performed by the studied projects. Note that the sum is greater than 40 since a project can leverage more than one model, each with a different post-processing strategy.

\textbf{Output Structuring} is the most frequently identified post-processing step. This process typically employs regular expressions to resolve inconsistencies when prompts omit format requirements or when the model fails to fulfill them at inference time. 

\begin{table}[t]
\footnotesize
\centering
\caption{Model terms of use violations and project counts}
\label{tab:term_of_use_violation_models}
\resizebox{\columnwidth}{!}{
\begin{tabular}{p{1.1cm} p{6cm} r r}
\toprule
\textbf{Models (Version)} & \textbf{ToU Potential Violations} & \textbf{Ref.} & \textbf{\# Prj}\\
\midrule
GPT\newline(3.5, 4o, 4o-mini, 4) &
``automation of high-stakes decisions in sensitive areas without human review'' & \cite{openaitou2025} & 14\\ \hline
Gemini\newline(1.5, 2.0, 2.5 Pro) &``Don't rely on the Services for medical, mental health, legal, financial, or other professional advice'' & \cite{geminitos2025} & 5\\ \hline
Gemma\newline(2-9B) & ``Perform or facilitate dangerous, illegal, or malicious activities, including [..] Engagement in the illegal or unlicensed practice of any vocation or profession, including, but not limited to, legal, medical, accounting, or financial professional practices.''& \cite{gemmatos2025} & 2\\ \hline
LLaMA\newline(2, 3.2) & ``Engage in the unauthorized or unlicensed practice of any profession, including, but not limited to, financial, legal, medical/health, or related professional practices.'' & \cite{metatos2025} & 6
\\ \hline
Qwen\newline(2, 2.5) &
``Do not perform or facilitate the activities [..] such as providing tailored legal, medical/health, or financial advice, making automated decisions in domains that affect an individual’s rights[..]''
& \cite{qwentos2025} & 4 \\ \hline
Deepseek\newline(v3-0324) &``You should not treat the Outputs as professional advice. Specifically, when using this service to consult on medical, legal, financial, or other professional issues[..]'' & \cite{dstos2025} & 1 \\
\bottomrule
\end{tabular}
}
\vspace{-4mm}
\end{table}

The \textbf{Output Consistency} post-processing step applies expert-defined rules to ensure the model outputs adhere to domain-specific constraints. 
For example, in autonomous driving, these rules ensure that a recommended acceleration is within physically plausible limits. In the medical domain, they prevent the simultaneous prescription of incompatible drugs, while in finance, they enforce risk management by limiting the investment threshold for a single stock. 

The \textbf{Auto-Refiner} category includes projects that feature system-level feedback loops evaluating the ML output and re-invoke models, rather than simply leveraging the ML model chain-of-thought reasoning. For instance, a trading strategy is evaluated using historical data, adjusted based on the findings, and retested to improve performance. Similarly, a drone route is continually optimized to enhance both time and energy efficiency.

The \textbf{Multi-Agent Coordination} category includes projects that rely on multiple ML models consulting and selecting a final answer based on their agreement. This approach is used, for example, in medical areas to obtain a diagnosis.


Finally, the \textbf{Confidence Threshold} category
restricts decision-making to outputs that exceed a predefined confidence level. While LLMs often rely on perplexity for this purpose, traditional ML models use inference probabilities. For instance, a phishing detector might only block an email if the probability of it being malicious exceeds 80\%. Similarly, a drone detection system may only trigger an alert once it reaches a 90\% confidence threshold.\\

\textbf{RQ$_3$:} \textbf{\rqfour}


Based on our sampling strategy, all 45 projects under analysis fall within areas classified as ``High-Risk'' by the EU AI Act. In this category, human-in-the-loop is required by the Act.
Truly, this can, in principle, be achieved by introducing human vigilance when using the system, but this is not explicitly implemented. 

Regarding potential ToU violations, among 45 projects analyzed, 25 (about 56\%) use an ML model in violation of its ToU.
\tabref{tab:term_of_use_violation_models} lists the ML models involved with the quotations from their ToU, grouped by provider when sharing the same regulatory conditions, and the number of projects using a model from that family violating its ToU. Note that the sum is greater than 25 since a project may have more than one model.  
All potential ToU violations involve proprietary LLMs, whereas we do not observe a comparable issue for open-weight models. However, this mostly happens because, based on our analysis, ToU are mostly available in popular proprietary LLMs. 
As shown in \tabref{tab:term_of_use_violation_models}, GPT-4o and Gemini models appear to be the most used when a potential ToU violation is raised. Furthermore, ToU potential violations may mostly occur either because the ML model is used to enact automation when it was not conceived to do so, or because it is used in a domain---\eg healthcare or finance---where it should not be used.

%% file: threats.tex
Threats to \textbf{construct validity} concern the relationship between theory and observation. The main threats are related to how we are operationalizing the identification of (i) automated (ML-based) decisions and (ii) their post-processing. To make the work repeatable, we have detailed the information we looked for in Section \ref{sec:design} and, secondarily, in our replication package~\cite{replication}.

Threats to \textbf{internal validity} concern factors internal to our study that can influence our findings. In general, such threats are limited in studies like ours, as we do not make any cause-and-effect claim. At the same time, we are aware that there may be imprecision and errors in our analyses. We mitigated this by having one author perform the initial coding and a second author iterate over it.

Threats to \textbf{external validity} concern the generalizability of our findings. Our study analyzes a few (173, of which 45 make fully automated decisions) open-source projects, selected as explained in Section \ref{sec:design}, using ML models. While the findings may not generalize to many real-world, commercial applications, they (i) show that the problem is there, and requires attention, and (ii) help to define a more general compliance assessment framework.

%% file: conclusion.tex
In this paper, we performed a preliminary analysis on how ML models are used in open-source projects for decision-making. We analyzed 173 high-risk projects hosted on GitHub that use either open-weight models from Hugging Face or closed models such as GPT or Gemini. 

Out of 173 analyzed projects, 45 use ML models for automated decision-making, performing a wide range of actions beyond recommendations, including automated control, security threat detection, and decision optimization. 
Most of the studied projects (40) do not depend directly on raw ML outputs but perform additional post-processing steps before using them, aiming to mitigate potential threats arising from relying solely on ML outputs. At the same time, however, we noticed that in most cases, the ML model's use may not be fully compliant with AI regulations such as the EU AI Act~\cite{aiact2024} and the model's specific terms of use.
In summary, the early results of this paper highlight how projects use ML models to make ``high-risk" decisions, in relation to regulations and terms of use. 

Future work will extend the preliminary analyses we performed and develop approaches and tools to automatically identify and verify regulatory compliance of ML model usage in software systems.